\documentclass[apj]{aastex63}
\usepackage{lineno}
\usepackage{gensymb}
\usepackage{lineno}
\usepackage{amsmath}
\usepackage{wasysym}
\usepackage{float}
\usepackage{multirow}
 \usepackage{graphicx}
\usepackage{natbib}

\graphicspath{{./}{figures/}}
\NewPageAfterKeywords
\begin{document}
\shorttitle{Multi-component adiabat}
\shortauthors{Graham, Lichtenberg, Boukrouche, \& Pierrehumbert}
\title{A multispecies pseudoadiabat for simulating condensable-rich exoplanet atmospheres}

\author[0000-0001-9289-4416]{R.J. Graham}
\affil{Atmospheric, Oceanic and Planetary Physics, Department of Physics, University of Oxford, UK}
\author[0000-0002-3286-7683]{Tim Lichtenberg}
\affil{Atmospheric, Oceanic and Planetary Physics, Department of Physics, University of Oxford, UK}
\author[0000-0002-5728-5129]{Ryan Boukrouche}
\affil{Atmospheric, Oceanic and Planetary Physics, Department of Physics, University of Oxford, UK}
\author[0000-0002-5887-1197]{Raymond T. Pierrehumbert}
\affil{Atmospheric, Oceanic and Planetary Physics, Department of Physics, University of Oxford, UK}

\correspondingauthor{R.J. Graham}
\email{robert.graham@physics.ox.ac.uk}


\begin{abstract}
Central stages in the evolution of rocky, potentially habitable planets may play out under atmospheric conditions with a large inventory of non-dilute condensable components. Variations in condensate retention and accompanying changes in local lapse rate may substantially affect planetary climate and surface conditions, but there is currently no general theory to effectively describe such atmospheres. In this article, expanding on the work by \citet{li2018moist}, we generalize the single-component moist pseudoadiabat derivation in \citet{Pierrehumbert:2010-book} to allow for multiple condensing components of arbitrary diluteness and retained condensate fraction. The introduction of a freely tunable retained condensate fraction allows for a flexible, self-consistent treatment of atmospheres with non-dilute condensable components. To test the pseudoadiabat's capabilities for simulating a diverse range of climates, we apply the formula to planetary atmospheres with compositions, surface pressures, and temperatures representing important stages with condensable-rich atmospheres in the evolution of terrestrial planets: a magma ocean planet in a runaway greenhouse state; a post-impact, late veneer-analogue planet with a complex atmospheric composition; and an Archean Earth-like planet near the outer edge of the classical circumstellar habitable zone. We find that variations in the retention of multiple non-dilute condensable species can significantly affect the lapse rate and in turn outgoing radiation and the spectral signatures of planetary atmospheres. The presented formulation allows for a more comprehensive treatment of the climate evolution of rocky exoplanets and early Earth analogues.
\end{abstract}
\section{Introduction}

The vertical temperature structure of an atmosphere is a primary determinant of planetary climate and surface evolution. The temperature gradient in convecting regions of atmospheres can generally be described by some form of adiabat (in the reversible case) or pseudoadiabat (in the irreversible case) that models the expansion and cooling of parcels of air as they are buoyed upward. In the case where none of the gases in the atmosphere can condense at the local temperature and pressure, the atmosphere follows a ``dry adiabat,'' where the temperature profile is determined entirely by the cooling of a convecting parcel as it expands and does work on its surroundings. In the case where the atmosphere is composed of a dry component mixed with ``wet'' components that can condense under the local temperature/pressure conditions, the latent heat released by condensation partially offsets the cooling by expansion work, leading to a ``moist adiabat'' with a lapse rate that is less steep than would be the case under dry conditions. 

Moist adiabats and pseudoadiabats with one \citep[e.g.][]{ingersoll69,Pierrehumbert:2010-book} or recently multiple \citep{li2018moist} condensables have been derived and used to investigate planetary atmospheres with a range of compositions. \citet{ingersoll69} exploited entropy conservation to derive both an adiabat (with all condensate retained in the column) and a pseudoadiabat (with instantaneous rainout of all condensate) to describe atmospheres with a single non-dilute condensable component in the context of modeling Venus undergoing a runaway greenhouse effect. \citet{li2018moist} took the same entropy-based approach to derive a fully reversible, condensate-retaining moist adiabat with multiple condensing components to describe the atmospheres of gas giant planets. An alternate approach to the derivation of moist (pseudo)adiabats is to begin with a statement of energy conservation. \citet{weidenschilling1973atmospheric} used energy conservation to derive a single-component moist pseudoadiabat to model the Jovian planets, and \citet{Pierrehumbert:2010-book} took the same approach during a pedogogical discussion of atmospheric thermodynamics. 

Here, we apply the energy conservation approach to derive a lapse rate formula that allows for multiple condensing components and the specification of arbitrary retained condensate fractions. For atmospheres with dilute condensable components, regardless of retained condensate fraction, the temperature-pressure profiles produced by this formula differ negligibly from those produced by the fully reversible adiabat derived in \citet{li2018moist}. However, under non-dilute conditions, the retained condensate fraction controls the effective specific heat of the atmosphere, which in turn strongly influences the lapse rate of the atmosphere. With the assumption of complete condensate retention that is made in \citet{li2018moist}, non-dilute atmospheres develop upper atmospheres composed almost entirely of condensate, with gas at times making up less than $1\%$ of moles at low pressures/temperatures where nearly all of the gas has condensed. When the ratio of gas to condensate is very small, a change in the temperature of a mole of the gas also requires changing the temperature of many moles of a solid/liquid phase, effectively increasing the specific heat capacity of the gas and making its temperature less sensitive to expansion work done on the surroundings. It is unclear how realistic that situation might be, so it is important to be able  to vary the degree of condensate retention when modeling exotic atmospheres with non-dilute condensable components. 

We derived the following lapse rate equation to deal with this issue explicitly:

\begin{linenomath}
\large
\begin{align}
\label{eqn:adiabat}
    \frac{d\ln T}{d\ln P}&=\frac{x_d+\sum\limits_ix_{v,i}}{x_d\frac{c_{d}x_d+\sum\limits_i\left(x_{v,i}(c_{v,i}-R\beta_i+R\beta_i^2)+\alpha_ix_{c,i}c_{c,i}\right)}{R\left(x_d+\sum\limits_i\beta_ix_{v,i}\right)}+\sum\limits_i\beta_ix_{v,i}}
\end{align}
\end{linenomath}
where $\frac{d\ln T}{d\ln P}$ is the lapse rate of temperature $T$ with respect to total pressure $P$; $x_{\text{\{\}}}$ is the mole fraction of either a dry gas component (or gas mixture) ($x_{d}$), a condensable vapor component ($x_{v,i}$), or a condensate component ($x_{c,i}$); $c_{\text{\{\}}}$ is the molar specific heat at constant pressure [J K$^{-1}$ mol$^{-1}$] of a dry gas/mixture ($c_d$), a condensing gas ($c_{v,i})$ or a condensate ($c_{c,i}$); $R=8.314$ [J K$^{-1}$ mol$^{-1}$] is the universal ideal gas constant; $\beta_{i}=\frac{L}{RT}$ where $L$ [J mol$^{-1}$] is the latent heat of condensable component $i$; and $\alpha_i$ is the mole fraction of condensate that is retained in the column instead of raining out for species $i$. It is important to note that the mole fractions $x_{\text{\{\}}}$ in this equation obey the relation $x_d + \sum\limits_i x_{v,i} + \sum\limits_i x_{c,i}$ = 1, and that $x_{c,i}$ is the total condensate produced in a given parcel of air during its ascent from the surface, including both retained ($\alpha_ix_{c,i}$) and rained-out ($(1-\alpha_{i})x_{c,i}$) components. The mole fractions of components \textit{in the atmosphere} can be calculated with $X^i_{\text{\{\}}} = \frac{x_\text{\{\}}}{1-(1-\alpha_i)x_{c,i}}$ for the gaseous components and, for the condensate components, $X^i_c = \frac{\alpha_i x_{\rm c,i}}{1-(1-\alpha_i)x_{c,i}}$.

This pseudoadiabat can be applied to model a very broad range of atmospheric compositions. To illustrate this, we use the formula to explore several scenarios where condensable components may dominate terrestrial atmospheres:  (i) a magma ocean planet with a partially molten surface, (ii) an impact-induced, transiently-reducing atmosphere during the late veneer, and (iii) the atmosphere of an Earth-like planet in the early stages of its habitable phase, with high CO$_2$ partial pressure and temperate surface climate. Each case has at least one abundant condensable species, so the degree of condensate retention is an especially important consideration in calculating temperature structure of these atmospheres. Previous work on atmospheres with non-dilute condensable components has demonstrated that these planets exhibit globally weak temperature gradients regardless of spin state \citep{pierrehumbert2016dynamics,ding2018global}, suggesting one-dimensional atmospheric column models are a useful tool for exploring the properties of these atmospheres.

The manuscript is structured as follows. In Sect. \ref{sec:adiabat} we derive the adiabat formula. Readers mainly interested in the main features of the adiabat, and less in the derivation itself, may continue in Sect. \ref{sec:features} ff., where we illustratively point out and discuss the influence on the lapse rate and condensation. In Sect. \ref{sec:cases} we demonstrate the impact of the retained condensate fraction on atmospheric structure and radiative transport for the three reference scenarios mentioned in the previous paragraph. In Sect. \ref{sec:discussion} we discuss the relevance for exoplanet atmospheres and the near-surface evolution of prebiotic Earth and highlight some caveats with our analysis. We summarise and conclude in Sect. \ref{sec:conclusions}.

\section{Multi-component pseudoadiabat}  \label{sec:adiabat}
\subsection{Derivation} \label{sec:derivation}
In this section, we derive the formula for the multi-condensable moist pseudoadiabat from basic thermodynamic principles. Readers with less interest in the mathematical derivation may skip to Section \ref{sec:features} to see the application of the formula to condensable-rich atmospheres. This derivation is a generalization of the single-component pseudoadiabat derivation presented in Section 2.7.2 of \citet{Pierrehumbert:2010-book}. We begin with a statement of the first law of thermodynamics for a parcel of gas with $n_d$ moles of non-condensable substance, $\sum\limits_i n_{v,i}$ moles of vapor of condensable substance, and $\sum\limits_i \alpha_i n_{c,i}$ moles of condensate, where $n_{c,i}$ is the total number of moles of substance $i$ that have condensed over the history of the parcel and $\alpha_i\in$ [0,1] is the fraction of condensate $i$ that has been retained in the parcel. The subscript $i$ iterates over the number of condensable species, and the subscript $d$ represents the non-condensable (``dry'') phase, which may actually be the molar mean of arbitrarily many dry phases. If temperature, non-condensable pressure, and condensable pressure are changed respectively by infinitesimal increments $dT$, $dP_d$, and $\sum\limits_i dP_i$, enthalpy conservation for the parcel can be stated as:
\begin{linenomath}
\begin{align}
\begin{split}
    &\left(n_d+\sum\limits_i (n_{v,i} + \alpha_i n_{c,i})\right)\delta Q =\\ 
    &n_d c_{d} dT - V_ddP_d+ \sum\limits_i \left(n_{v,i} c_{v,i} dT + \alpha_i n_{c,i}c_{c,i}dT - V_idP_i +L_idn_{v,i}\right)
    \end{split}
\end{align}
\end{linenomath}
\begin{linenomath}
\begin{align}
\begin{split}
&\left(n_d+\sum\limits_i (n_{v,i} + \alpha_i n_{c,i})\right)\delta Q=\\
    &n_d c_{d} dT - n_dRT\frac{dP_d}{P_d}+ \sum\limits_i \left(n_{v,i} c_{v,i} dT + \alpha_i n_{c,i}c_{c,i}dT - n_{v,i}RT\frac{dP_i}{P_i} +L_idn_{v,i}\right)
\end{split}
\end{align}
\end{linenomath}
where $\delta Q$ is the change in total energy per mole of the parcel ($=0$ if the parcel is an energetically closed system, as we will assume), the $-VdP$ terms represent the work done to change pressures, the second line comes from the ideal gas law ($V=\frac{nRT}{P}$), $R$ is the universal gas constant [J K$^{-1}$ mol$^{-1}$],  $L_i$ is the molar latent heat for each condensable species [J mol$^{-1}$], and $dn_{v,i}$ represents the change in number of moles of each condensable substance due to condensation.  

Next, we set $\delta Q=0$, divide the equation by $n_dT$, and define $\eta_{v,i}=\frac{n_{v,i}}{n_d}$:
\begin{linenomath}
\begin{align}
    0&=c_{d} \frac{dT}{T} - R\frac{dP_d}{P_d}+ \sum\limits_i \left(\frac{n_{v,i}}{n_d} c_{v,i} \frac{dT}{T} + \alpha_i \frac{n_{c,i}}{n_d}c_{c,i}dT - n_{v,i} - \frac{n_{v,i}}{n_d}R\frac{dP_i}{P_i} +\frac{L_idn_{v,i}}{n_dT}\right)\\
    &=c_{d} \frac{dT}{T} - R\frac{dP_d}{P_d}+ \sum\limits_i \left(\eta_{v,i} c_{v,i} \frac{dT}{T} + \alpha_i\eta_{c,i} c_{c,i} \frac{dT}{T} - \eta_{v,i}R\frac{dP_i}{P_i} +\frac{L_i}{T}d\eta_{v,i}\right)
\end{align}
\end{linenomath}
where $\eta_{\{\},i}$ is the molar mixing ratio of condensable or condensate species $i$ with the non-condensable gas. 

Next, we rewrite $d\eta_{v,i}$ in terms of changes to pressure and make use of the fact that the condensable pressures follow the ideal-gas form of Clausius-Clapeyron equation (see eqn. \ref{eqn:clausius}) to rewrite $\frac{dP_i}{P_i}$ in terms of a change to the temperature (We note that application of this form of the Clausius-Clapeyron equation implies the assumption that the specific volume of the condensate phase being described is negligible compared to that of the vapor phase of the same species--this assumption begins to break down near the critical point where gas and liquid phases become indistinguishable, so in some high temperature and/or pressure atmospheres, the accuracy of the pseudoadiabat will be reduced):

\begin{linenomath}
\begin{align}
d\eta_{v,i}&=d\left(\frac{P_i}{P_d}\right)\\
&=\frac{P_i}{P_d}d\ln\left(\frac{P_i}{P_d}\right)\\
&=\eta_{v,i}\left(d\ln(P_i)-d\ln(P_d)\right)
\end{align}
\end{linenomath}
and
\begin{linenomath}
\begin{align}
P_i&=P_{sat,i}(T)\\ \frac{dP_i}{P_i}&=d\ln{P_i}=\frac{d\ln(P_{sat,i}(T))}{dT}dT\\
&=\frac{L_i}{RT^2}dT\label{eqn:clausius}\\
&=\frac{L_i}{RT}d\ln{T}
\end{align}
\end{linenomath}
both of which can now be substituted into the statement of the first law:
\begin{linenomath}
\begin{align}
\begin{split}
    &0 = c_{d}d\ln{T}-Rd\ln{P_d} +...\\
    &\sum\limits_i\left(\eta_{v,i}c_{v,i}d\ln{T}+ \alpha_i\eta_{c,i} c_{c,i}d\ln{T}-\eta_{v,i}\frac{L_i}{T}d\ln{T}+\frac{\eta_{v,i}L_i}{T}\left(\frac{L_i}{RT}d\ln{T}-d\ln{P_d}\right)\right)
\end{split}
\end{align}
\end{linenomath}
which now only has differentials of $\ln{P_d}$ and $\ln{T}$. We can now gather terms with each differential:
\begin{linenomath}
\begin{align}
\begin{split}
  &d\ln{P_d}\left(R+\sum\limits_i\frac{\eta_{v,i}L_i}{T}\right)=\\
  &d\ln{T}\left(c_{d}+\sum\limits_i\left(\eta_{v,i}\left(c_{v,i}-\frac{L_i}{T}+\frac{L_i^2}{RT^2}\right) + \alpha_i\eta_{c,i}c_{c,i}\right)\right)
\end{split}
\end{align}
\end{linenomath}
and solve for  $\frac{d\ln{T}}{d\ln{P_d}}$:
\begin{linenomath}
\begin{equation}
    \frac{d\ln{T}}{d\ln{P_d}} = \frac{R+\sum\limits_i\frac{\eta_{v,i}L_i}{T}}{c_{d}+\sum\limits_i\left(\eta_{v,i}\left(c_{v,i}-\frac{L_i}{T}+\frac{L_i^2}{RT^2}\right)+\alpha_i\eta_{c,i}c_{c,i}\right)}
\end{equation}
\end{linenomath}
which represents the lapse rate with respect to changes in the noncondensable component's pressure. 

Now, to get the lapse rate with respect to changes to the total pressure, we note:
\begin{linenomath}
\begin{equation}
    \frac{d\ln{T}}{d\ln P} = \frac{d\ln T}{d\ln P_d}\frac{d\ln P_d}{d\ln P}
\end{equation}
\end{linenomath}
where $P=P_d+\sum\limits_iP_i$. 

So, the next step is to write $\frac{d\ln P_d}{d\ln P}$ in terms of convenient variables:
\begin{linenomath}
\begin{align}
    \frac{d\ln P_d}{d\ln P}&=\frac{P}{P_d}\frac{dP_d}{dP}\\
    &=\frac{P_d+\sum\limits_iP_i}{P_d}\left(\frac{dP_d}{d\left(P_d+\sum\limits_iP_i\right)}\right)\\
    &=\left(1+\sum\limits_i\frac{P_i}{P_d}\right)\left(1+\sum\limits_i\frac{dP_i}{dP_d}\right)^{-1}\\
    &=\left(1+\sum\limits_i\frac{P_i}{P_d}\right)\left(1+\sum\limits_i\left(\frac{dP_i}{dT}/\frac{dP_d}{dT} \right)\right)^{-1}\\
    &=\left(1+\sum\limits_i\eta_{v,i}\right)\left(1+\sum\limits_i\left(\frac{L_iP_i}{RT^2}\right)\left(\frac{T}{P_d}\frac{d\ln T}{d\ln P_d}\right)\right)^{-1}\\
    &=\frac{1+\sum\limits_i\eta_{v,i}}{1+\sum\limits_i\frac{L_i}{RT}\eta_{v,i}\frac{d\ln T}{d\ln P_d}}\\
    &=\frac{x_d+\sum\limits_ix_{v,i}}{x_d+\sum\limits_i\beta_ix_{v,i}\frac{d\ln T}{d\ln P_d}}
\end{align}
\end{linenomath}
where $\beta\equiv\frac{L_i}{RT}$ and $x_{\{\}}$ represents the mole fraction of a substance, such that $\frac{x_{v,i}}{x_d}=\eta_{v,i}$. 

Now we can write down $\frac{d\ln T}{d\ln P}$. We substitute $\frac{x_{v,i}}{x_d}$ for $\eta_{v,i}$ and multiply $x_d$ from the denominator to avoid dealing with $\eta_{v,i}$ going to infinity at vanishing non-condensable partial pressure, and substitute $\beta_i$ for $\frac{L_i}{RT}$ for compactness:
\begin{linenomath}
\begin{align}
    \frac{d\ln T}{d\ln P}&=\frac{d\ln P_d}{d\ln P}\frac{d\ln T}{d\ln P_d}\\
    &=\frac{x_d+\sum\limits_ix_{v,i}}{x_d+\sum\limits_i\beta_ix_{v,i}\frac{d\ln T}{d\ln P_d}}\frac{d\ln T}{d\ln P_d}\\
    &=\frac{x_d+\sum\limits_ix_{v,i}}{x_d\left(\frac{d\ln T}{d\ln P_d}\right)^{-1}+\sum\limits_i\beta_ix_{v,i}}\\
    &=\frac{x_d+\sum\limits_ix_{v,i}}{x_d\frac{c_{d}x_d+\sum\limits_i\left(x_{v,i}(c_{v,i}-R\beta_i+R\beta_i^2)+\alpha_ix_{c,i}c_{c,i}\right)}{R\left(x_d+\sum\limits_i\beta_ix_{v,i}\right)}+\sum\limits_i\beta_ix_{v,i}}
\end{align}
\end{linenomath}
which we later demonstrate to be equivalent to the reversible multi-condensable adiabat derived in \citet{li2018moist}, with the fraction of retained condensate in that formula set equal to $\alpha$ instead of one (see Section \ref{app:1}).

It is simple to show that this formula yields reasonable answers at various limits. For example, the dry adiabat can be derived by setting $\sum\limits_ix_{v,i}=\sum\limits_ix_{c,i}=0$ and $x_d=1$ (e.g. assuming a column of non-condensable gas):
\begin{linenomath}
\begin{align}
    \frac{d\ln T}{d\ln P}\bigg|_{dry} &=\frac{1+\sum\limits_i0}{1\times\frac{c_{d}\times1+\sum\limits_i\big(0\times(c_{v,i}-R\beta_i+R\beta_i^2)+\alpha_i\times 0\times c_{c,i}\big)}{R\left(1+\sum\limits_i\beta_i\times0\right)}+\sum\limits_i\beta_i\times0}\\
    &=\frac{R}{c_{d}}
\end{align}
\end{linenomath}
which is the correct formula. Similarly, we can set $x_d=\sum\limits_i\alpha_ix_{c,i}=0$ to represent an atmosphere made entirely of condensing species with instantaneous rainout of condensates:
\begin{linenomath}
\begin{align}
    \frac{d\ln T}{d\ln P}\bigg|_{moist}&=\frac{\sum\limits_ix_{v,i}}{\sum\limits_i\beta_ix_{v,i}}\\
    &=\frac{1}{\sum\limits_i\beta_ix_{v,i}}
\end{align}
\end{linenomath}
which yields the familiar Clausius Clapeyron equation when evaluated with a single component: $\frac{d\ln T}{d\ln P}\big|_{CC}=\frac{1}{\beta}=\frac{RT}{L}$.

\subsection{Convergence to reversible multi-condensable adiabat}\label{app:1}

Here we show that the multi-condensable adiabat derived in \citet{li2018moist} can be obtained by setting the fraction of retained condensate to one in the formula derived here:
\begin{linenomath}
\begin{align}
     \frac{d\ln T}{d\ln P}&=\frac{x_d+\sum\limits_ix_{v,i}}{x_d\frac{c_{d}x_d+\sum\limits_i\big(x_{v,i}(c_{v,i}-R\beta_i+R\beta_i^2)+\alpha_ix_{c,i}c_{c,i}\big)}{R(x_d+\sum\limits_i\beta_ix_{v,i})}+\sum\limits_i\beta_ix_{v,i}}\\
     &=\frac{1+\sum\limits_i\eta_{v,i}}{\frac{c_{d}x_d+\sum\limits_ix_{v,i}\big(x_{v,i}(c_{v,i}-R\beta_i+R\beta_i^2)+\alpha_ix_{c,i}c_{c,i}\big)}{Rx_d(1+\sum\limits_i\beta_i\eta_{v,i})}+\sum\limits_i\beta_i\eta_{v,i}}\\
     &=\frac{1+\sum\limits_i\beta_i\eta_{v,i}}{\frac{c_{d}x_d+\sum\limits_i(x_{v,i}c_{v,i}+\alpha_ix_{c,i}c_{c,i})}{R(x_d+\sum\limits_ix_{v,i})}+\frac{\sum\limits_i(-\beta_i\eta_{v,i}+\beta_i^2\eta_{v,i})+\sum\limits_i\beta_i\eta_{v,i}+\left(\sum\limits_i\beta_i\eta_{v,i}\right)^2}{1+\sum\limits_i\eta_{v,i}}}\\
     &=\frac{1+\sum\limits_i\beta_i\eta_{v,i}}{\frac{c_{d}x_d+\sum\limits_i(x_{v,i}c_{v,i}+\alpha_ix_{c,i}c_{c,i})}{R(x_d+\sum\limits_ix_{v,i})}+\frac{\sum\limits_i(\beta_i^2\eta_{v,i})+(\sum\limits_i\beta_i\eta_{v,i})^2}{1+\sum\limits_i\eta_{v,i}}}\\
     &=\frac{1+\sum\limits_i\beta_i\eta_{v,i}}{\frac{\widehat{c_{p}}}{R}+\frac{\sum\limits_i(\beta_i^2\eta_{v,i})+\left(\sum\limits_i\beta_i\eta_{v,i}\right)^2}{1+\sum\limits_i\eta_{v,i}}\label{eqn:Li_reprod}}
\end{align}
\end{linenomath}
where $\widehat{c_{p}}=\frac{c_{d}x_d+\sum\limits_i(x_{v,i}c_{v,i}+\alpha_ix_{c,i}c_{c,i})}{\left(x_d+\sum\limits_ix_{v,i}\right)}$. This is identical to equation (1) in \citet{li2018moist}, except that the $\widehat{c_p}$ in their equation (1) has $\alpha_i=1$, representing complete condensate retention in the fully reversible limit. $\widehat{c_p}$ is the specific heat of the atmosphere per unit gaseous mole. Increases to the condensate loading of the atmosphere spread the energy required to change the temperature of the gas across a larger quantity of material without changing the amount of gas. This allows $\widehat{c_p}$ to exceed the specific heat capacity of either the gas or the condensate phases when the ratio of condensate to gas becomes large.

\subsection{Adiabat features} \label{sec:features}

\begin{figure}[tb]
    \centering
    \includegraphics[width=0.49\textwidth]{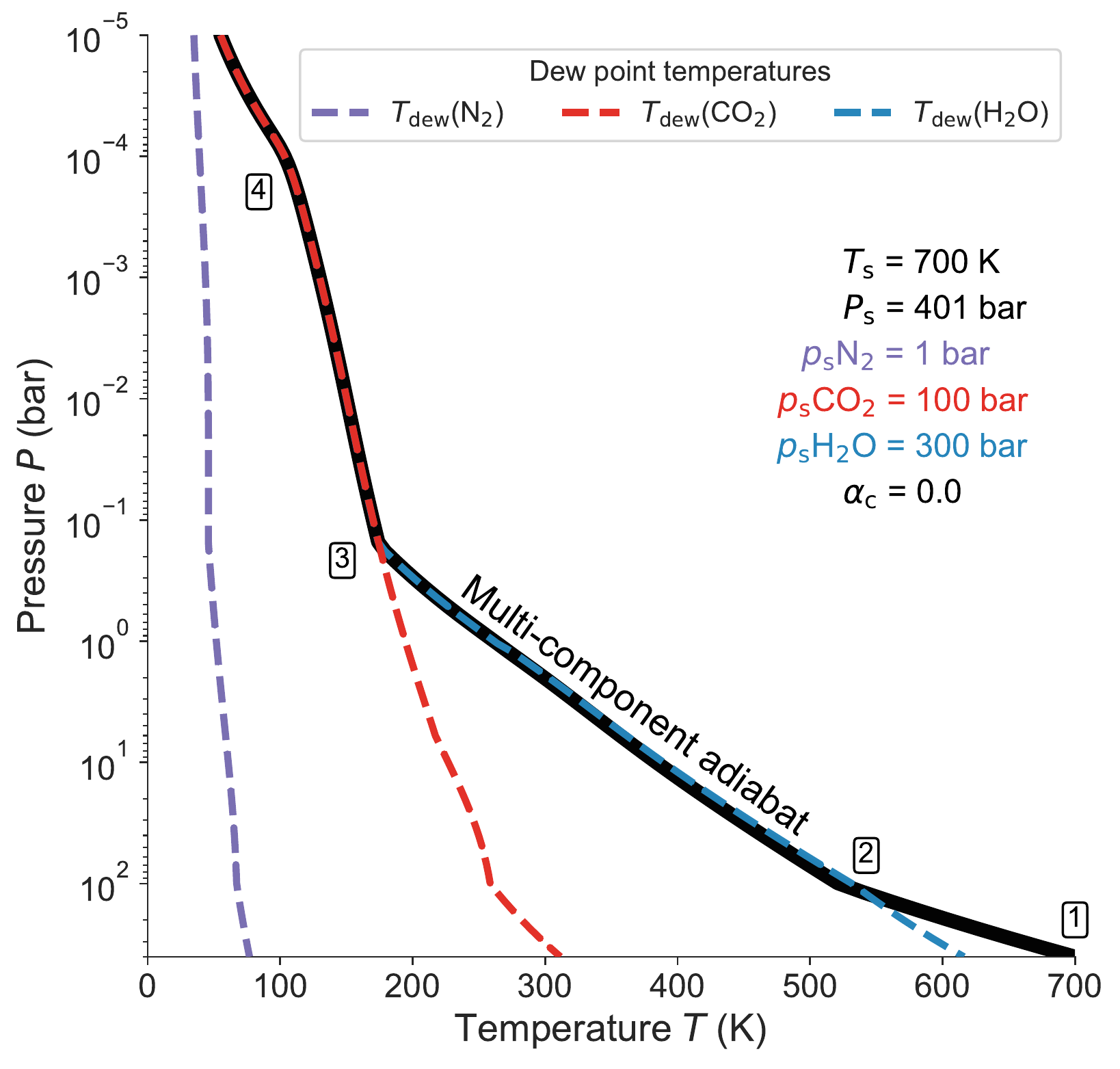}
    \caption{Annotated illustration of some basic adiabat features in a hypothetical N$_2$--CO$_2$--H$_2$O atmosphere with a surface temperature of 700 K. See main text for discussion.}
    \label{fig:adiabat_structure}
\end{figure}

In order to illustrate the behavior of an atmospheric column described by the pseudoadiabat, we discuss the major features in a hypothetical N$_2$-CO$_2$-H$_2$O atmosphere (Figure \ref{fig:adiabat_structure}). Point (1) is the surface. Because we chose subsaturated surface conditions for all three gases, here the atmosphere follows the dry adiabatic lapse rate and the composition of the atmosphere does not change with height/pressure. This situation continues until the atmosphere intersects the local dew-point temperature of H$_2$O ($T_{\rm dew}($H$_2$O); depicted by the blue curve) at point (2), with the dew-point temperature  approximated by \citep{Pierrehumbert:2010-book}:

\begin{linenomath}
\begin{align}
T_{\rm dew}(P) = \frac{T_0}{1-\frac{RT_0}{L}\ln{\frac{P}{P_{\rm sat}(T_0)}}}
\end{align}
\end{linenomath}
where $T_0$ is a reference temperature and the rest of the terms have been defined previously. For clarity, we note that this form of $T_{\rm dew}$ assumes that the latent heat of vaporization is constant. This assumption was not used in the derivation of the pseudoadiabat in this paper.

Point (2) is the ``cloud-base'' for H$_2$O; however, since we chose $\alpha$=0.0, all condensate instantaneously rains out, so technically there would be no clouds in this particular atmospheric profile. Above the cloud-base, latent heat release from H$_2$O's condensation decreases the lapse rate of the atmosphere by offsetting some of the cooling from expansion work.

Between points (2) and (3), the H$_2$O in the column rains out and becomes a minor constituent of the atmosphere, with CO$_2$ replacing it as the dominant component. Point (3) is the CO$_2$ cloud-base, where the pseudoadiabat crosses CO$_2$'s dew point temperature. In general, an abrupt change to the pseudoadiabatic lapse rate in the figures in this paper signals the initiation of condensation for some species. Because CO$_2$ is by far the dominant gas in the atmosphere at this point (as H$_2$O has rained out and N$_2$ is a minor background constituent), the pseudoadiabat essentially follows the CO$_2$ saturation vapor pressure curve above point (3) until it approaches point (4).

By point (4), most of the CO$_2$ in the atmosphere has rained out, leaving behind a tenuous upper atmosphere mostly consisting of N$_2$. The gradual bend in the profile that occurs at point (4) is due to a combination of the declining latent heat release as CO$_2$ becomes depleted and the smaller specific heat of N$_2$ ($\approx$29 J K$^{-1}$ mol$^{-1}$) compared to CO$_2$ ($\approx$36 J K$^{-1}$ mol$^{-1}$). A reduction in specific heat leads to a greater change in temperature for a given change in internal energy from expansion work done by the gas on its surroundings. 

\section{Case Studies}  \label{sec:cases}
\begin{table}[tb]
\centering
\begin{tabular}{llrrr}
Parameter & Unit & MO & LV & EE\\ \hline
$T_\mathrm{surf}$ & K & 1500 & 700 & 290 \\
$p \mathrm{H}_2\mathrm{O}$ & bar & 500 & 500 & 0.01 \\ 
$p \mathrm{CO}_2$ & bar & 100 & 48 & 29 \\ 
$p \mathrm{H}_2$ & bar & 100 & 8 & -- \\ 
$p \mathrm{N}_2$ & bar & 1 & 1 & 1 \\ 
$p \mathrm{CH}_4$ & bar & -- & 1 & -- \\ 
$p \mathrm{CO}$ & bar & -- & 0.05 & -- \\ 
$p \mathrm{NH}_3$ & bar & -- & 0.007 & -- \\ 
\end{tabular}
\caption{Case study reference settings. MO: Magma Ocean, LV: Late Veneer, EE: Exo-Earth}
\label{tab:cases}
\end{table}
\begin{figure*}[tbh]
    \centering
    \includegraphics[width=0.99\textwidth]{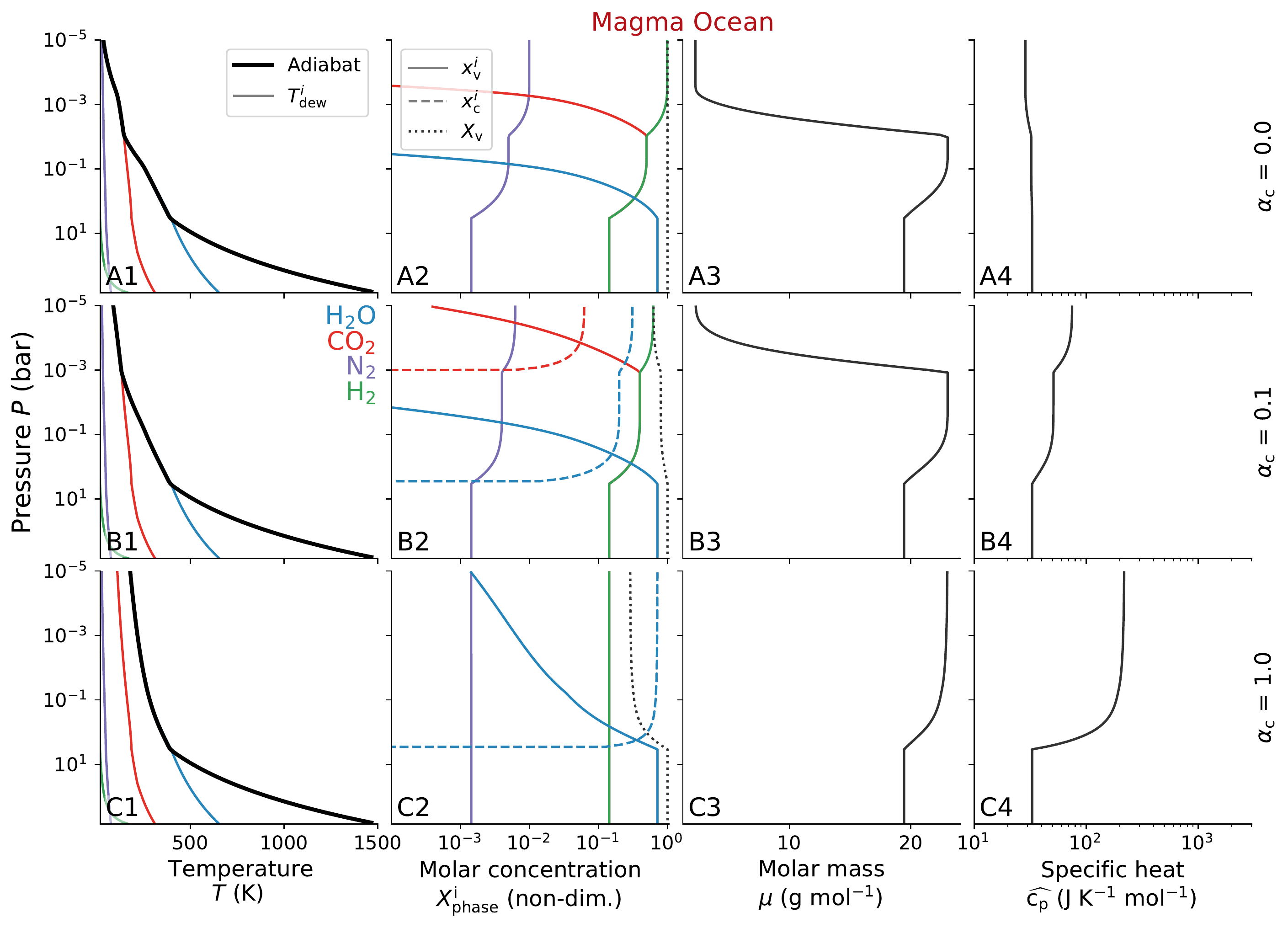}
    \caption{Magma ocean (MO) case study of a terrestrial planet with an inventory of about 1.85 Earth oceans (500 bar) H$_2$O, 100 bar CO$_2$, 100 bar H$_2$, and 1 bar N$_2$ in the atmosphere, with a surface temperature of 1500 K. The top row has retained condensate fraction $\alpha=0$, the middle row has $\alpha=0.1$, and the bottom row has $\alpha=1$. We note here that the $\alpha=1$ case, with full condensate retention, is unrealistic in atmospheres with nondilute condensable components and is included here as an illustration of the atmospheric profile produced by the condensate-retaining multicomponent moist adiabat derived in \citet{li2018moist}. The leftmost column shows the (pseudo)adiabat (thick black line) and the dew-point temperature of each component (thin colored lines), based on its partial pressure at a given atmospheric pressure. Varying the retained condensate fraction $\alpha$ from 0.0 to 0.1, and 1.0, substantially changes the lapse rate of the atmosphere and determines whether CO$_2$ condenses at all. The second column from the left displays the molar concentration of the different components of each atmosphere, including condensates when present. In the gaseous cases, $X^i_{\rm phase} = \frac{x_{\rm phase,i}}{1-(1-\alpha_i)x_{c,i}}$, while for the condensate, $X^i_c = \frac{\alpha_i x_{\rm c,i}}{1-(1-\alpha_i)x_{c,i}}$. The third column from the left displays the average molar mass of the gaseous components of the atmosphere as a function of pressure. The rightmost column displays the atmospheric specific heat per unit mole of gas, $\widehat{c_{p}}$ (defined below eqn. \ref{eqn:Li_reprod}), as a function of pressure.}
    \label{fig:MO}
\end{figure*}
\begin{figure*}[tbh]
    \centering
    \includegraphics[width=0.99\textwidth]{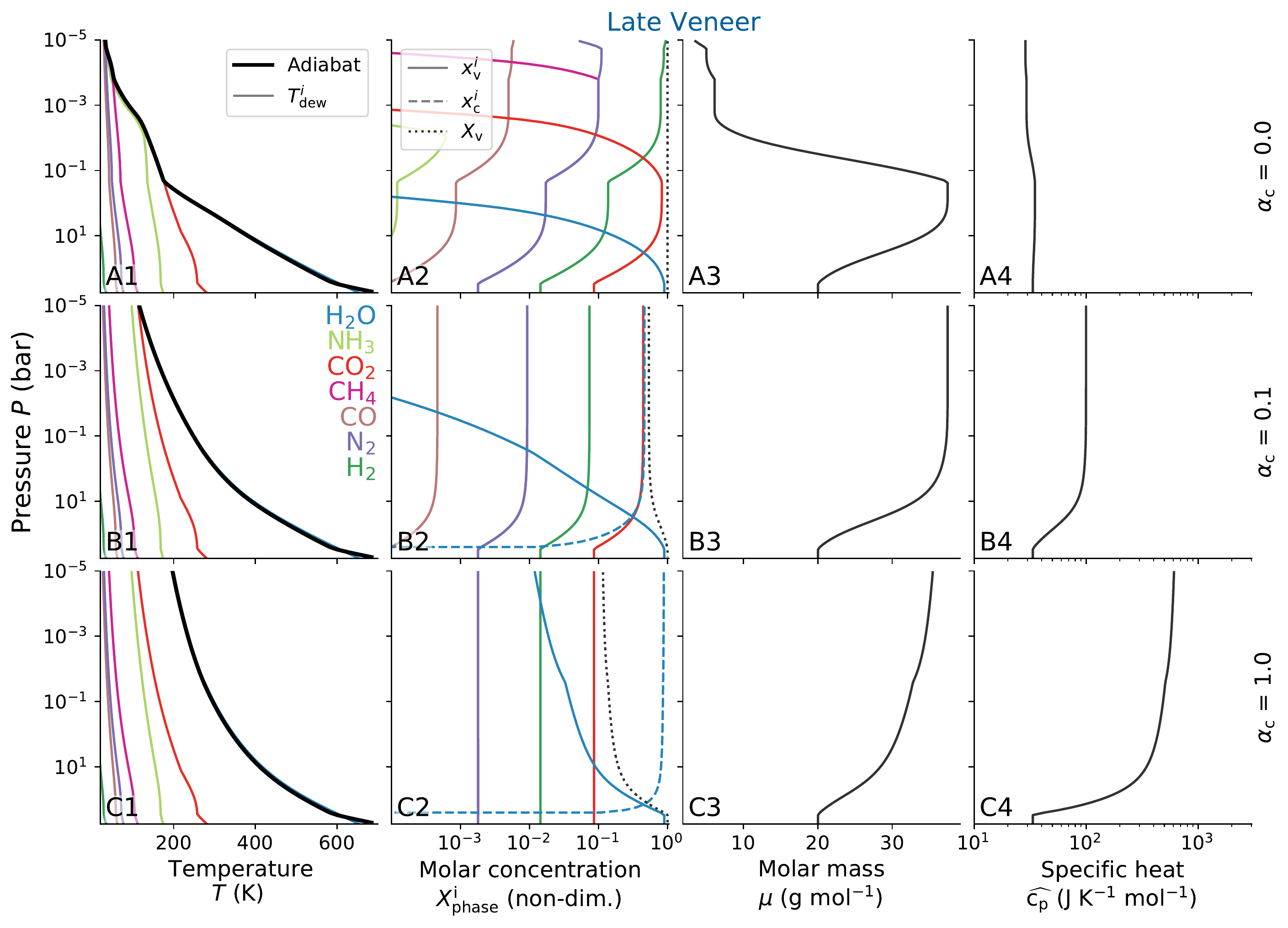}
    \caption{Late Veneer (LV) case study of a terrestrial planet after the impact of a reduced Vesta-sized body with an inventory of 500 bar H$_2$O, 48 bar CO$_2$, 8 bar H$_2$, 1 bar N$_2$, 1 bar CH$_4$, 0.05 bar CO, and 0.007 bar NH$_3$ in the atmosphere, with a surface temperature of 700 K. The top row has retained condensate fraction $\alpha=0$, the middle row has $\alpha=0.1$, and the bottom row has $\alpha=1$. The leftmost column shows (pseudo)adiabat (thick black line) and the dew-point temperature of each component (thin colored lines), based on its partial pressure at a given atmospheric pressure. Varying the retained condensate fraction $\alpha$ from 0.0 to 0.1, and 1.0, substantially changes the lapse rate of the atmosphere and determines whether CO$_2$, CH$_4$, and N$_2$ condense at all. The second column from the left displays the molar concentration of the different components of each atmosphere, including condensates when present. In the gaseous cases, $X^i_{\rm phase} = \frac{x_{\rm phase,i}}{1-(1-\alpha_i)x_{c,i}}$, while for the condensate, $X^i_c = \frac{\alpha_i x_{\rm c,i}}{1-(1-\alpha_i)x_{c,i}}$. The third column from the left displays the average molar mass of the gaseous components of the atmosphere as a function of pressure. The rightmost column displays the atmospheric specific heat per unit mole of gas, $\widehat{c_{p}}$ (defined below eqn. \ref{eqn:Li_reprod}), as a function of pressure.}
    \label{fig:LV}
\end{figure*}
\begin{figure*}[tbh]
    \centering
    \includegraphics[width=0.99\textwidth]{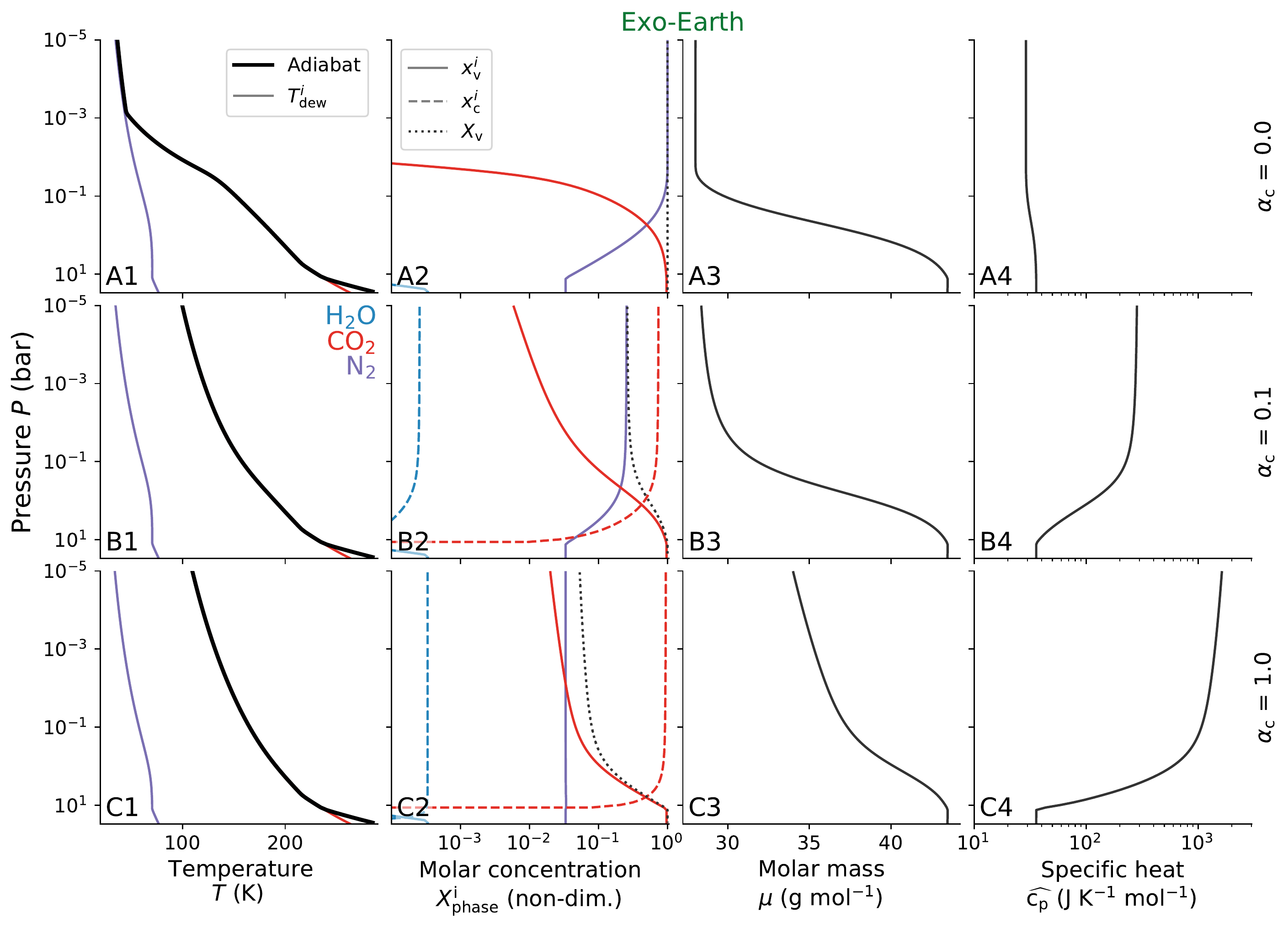}
    \caption{Exo-Earth (EE) case study of a terrestrial planet with 29 bar CO$_2$, 1 bar N$_2$, and 0.01 bar H$_2$O, with a surface temperature of 290 K. The top row has retained condensate fraction $\alpha=0$, the middle row has $\alpha=0.1$, and the bottom row has $\alpha=1$. The leftmost column shows the (pseudo)adiabat (thick black line) and the dew-point temperature of each component (thin colored lines), based on its partial pressure at a given atmospheric pressure. Varying the retained condensate fraction $\alpha$ from 0.0 to 0.1, and 1.0, substantially changes the lapse rate of the atmosphere and determines whether CO$_2$ condenses at all. The second column from the left displays the molar concentration of the different components of each atmosphere, including condensates when present. In the gaseous cases, $X^i_{\rm phase} = \frac{x_{\rm phase,i}}{1-(1-\alpha_i)x_{c,i}}$, while for the condensate, $X^i_c = \frac{\alpha_i x_{\rm c,i}}{1-(1-\alpha_i)x_{c,i}}$. The third column from the left displays the average molar mass of the gaseous components of the atmosphere as a function of pressure. The rightmost column displays the atmospheric specific heat per unit mole of gas, $\widehat{c_{p}}$ (defined below eqn. \ref{eqn:Li_reprod}), as a function of pressure.}
    \label{fig:OHZ}
\end{figure*}
\begin{figure*}[htb]
    \centering
    \includegraphics[width=0.99\textwidth]{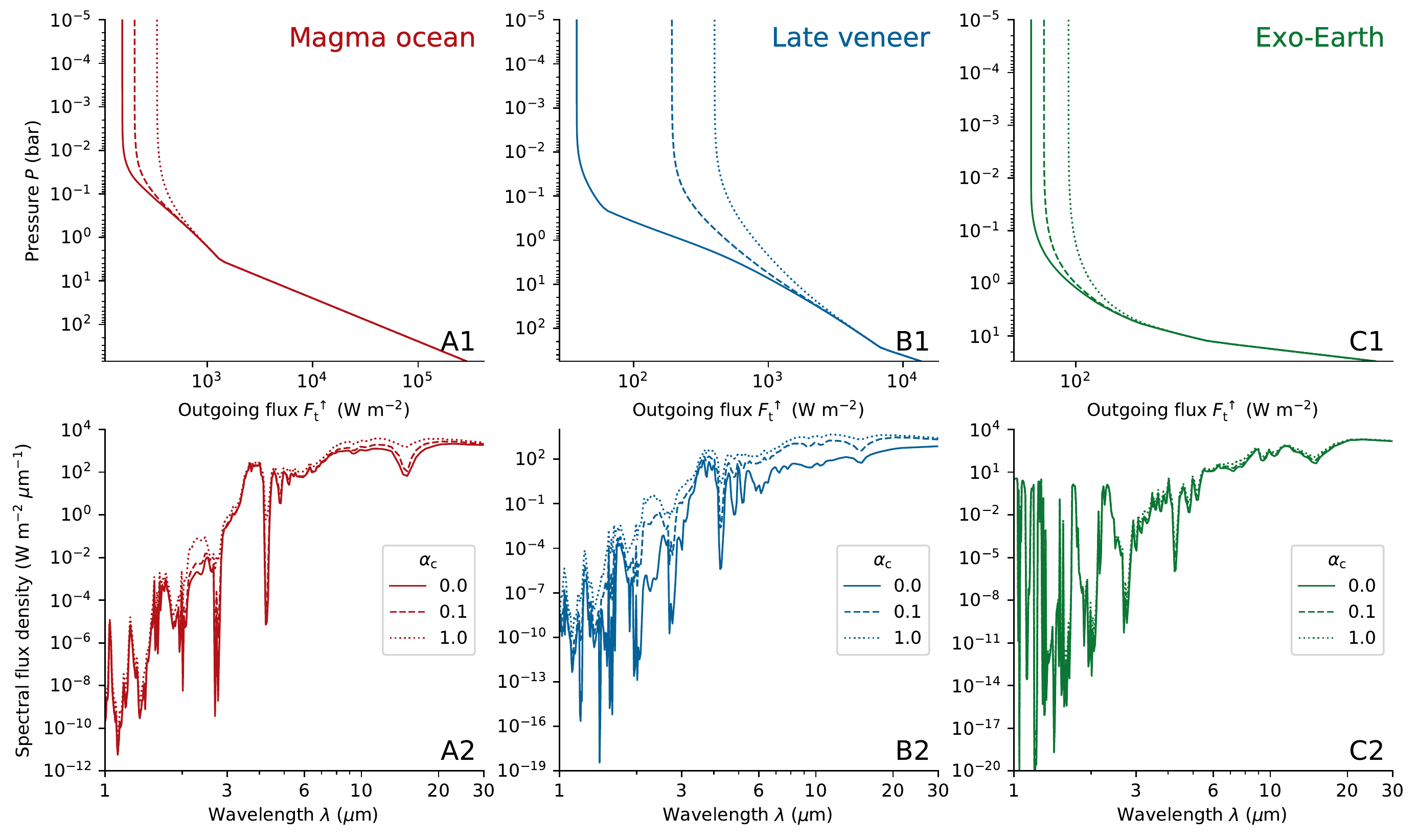}
    \caption{Deviations in planetary outgoing radiation and spectra for the three cases MO, LV, and EE, with varying condensate fraction $\alpha \in$ [0, 0.1, 1.0]. All three cases show substantial changes in their outgoing radiation with $\alpha$, with the LV case changing by more than an order of magnitude. The MO and LV $\alpha$ variations differ in their spectra particularly at near- and mid-infrared wavelengths, most pronounced between 2–3 $\mu$m, and between $\approx$10--20 $\mu$m for the MO case, and between $\approx$5--20 $\mu$m for the LV case. The EE $\alpha$ cases differ at $\approx$8 $\mu$m by up to one order of magnitude but are nearly  indistinguishable elsewhere.}
    \label{fig:RT}
\end{figure*}

In this section we use the newly-derived (pseudo)adiabat to model three non-dilute atmospheres chosen to represent snapshots from the first billion years of the evolution of a terrestrial planet, demonstrating the broad utility of the formula and the potential importance of retained condensate fraction to atmospheric behavior. For a given profile, we use a single choice of $\alpha$ for all atmospheric species at all levels of the atmosphere, but it is important to note that in reality $\alpha$ would be determined by an intricate interplay of cloud microphysics and large scale atmospheric dynamics, such that it would likely vary depending on the atmospheric constituent being considered and the position within the atmosphere. We use temperature-dependent specific heat capacities of gases and their liquid condensates from NIST \citep{lemmon2004nist}. We do not account for solid condensates in our calculations, but they would be treated identically to liquids. We use constant latent heat values from Table 2.1 in \citet{Pierrehumbert:2010-book}. 

In addition to simulating different atmospheric profiles and compositions, we assess their radiative and spectral properties using the \textsc{socrates} radiative transfer code \citep{edwards1996socrates}, solving the plane-parallel, two-stream approximated radiative transfer equation without scattering \citep[cf. the more extensive description in][]{LichtenbergJGRP21}. Opacity coefficients are tabulated and derived from the HITRAN database \citep{HITRAN2016,HITRAN_CIA1,HITRAN_CIA2}, making use of the line-by-line and collision-induced absorption coefficients for H$_2$O, CO$_2$, H$_2$ \citep{borysow2001high,borysow2002collision}, CH$_4$, N$_2$, and CO, and the H$_2$O continuum \citep{2012RSPTA.370.2520M}.
\subsection{Magma ocean Exo-Earth during planetary assembly phase}
Terrestrial planets are thought to pass through a magma ocean stage during their formation, with the planet's surface and interior rendered molten by the energy release from radionuclide decay and gravitational potential energy of the planet's constituent mass \citep[e.g.][]{ElkinsTanton2012AREPS}. The magma ocean period is crucial in determining the initial atmosphere and climate of a terrestrial planet, and the composition of the atmosphere overlying a magma ocean in turn exerts a primary control on the timescale of crystallization of the initially molten planetary mantle. Magma oceans can also form later in the evolution of a planet, for example after the planet goes through a runaway greenhouse event and accumulates a massive steam atmosphere. 

The magma ocean (MO) case demonstrates the control on cloud position exerted by the retained condensate fraction $\alpha$ via its influence on lapse rate (Fig. \ref{fig:MO}). H$_2$O makes up the lowest cloud deck, and the water cloud base occurs at the same point for all values of $\alpha$ (a pressure of several bars), since all three atmospheres follow the same dry adiabat in their respective non-condensing regions (subfigures A1, B1, C1). 

In the $\alpha=0.0$ case (Fig. \ref{fig:MO}), no condensate remains in the column, so the specific heat per unit mole of gas in the atmosphere remains in the range 30 to 40 J K$^{-1}$ mol$^{-1}$ throughout (subfigure A4). This low specific heat allows for rapid cooling as pressure decreases, particularly after the H$_2$O has largely been rained out and its latent heat effect has become insignificant at pressures below $\approx$10$^{-1}$ bar (subfigure A2). CO$_2$ then condenses at approximately 10$^{-2}$ bar and becomes an increasingly minor constituent of the atmosphere at lower and lower pressures, with H$_2$ becoming by far the dominant species at the top of the atmosphere.

The $\alpha=0.1$ case (Fig. \ref{fig:MO}) is similar to the full-rainout case, but not identical. With 10$\%$ of condensed moles retained, the effective specific heat of the atmosphere begins to increase above H$_2$O's cloud base. This reduces the atmospheric lapse rate and ultimately causes CO$_2$ to begin condensing at 10$^{-3}$ bar. In other words, the retention of 10$\%$ of condensate leads to an order of magnitude reduction in the cloud-base pressure of CO$_2$. At the top of the atmosphere, the effective atmospheric specific heat of the $\alpha=0.1$ case is approximately double that of the $\alpha=0.0$ case.

In the limit of full retention ($\alpha=1.0$, bottom row of Fig. \ref{fig:MO}), the effective specific heat increase due to condensate is more dramatic, with $c_p$ reaching a value of $\approx$200 J K$^{-1}$ mol$^{-1}$ (subfigure C4). With such large specific heat, the lapse rate above H$_2$O's cloud base is small enough that CO$_2$ never condenses, in contrast to the cases with smaller $\alpha$. This makes it clear that condensate retention in non-dilute atmospheres influences not just the position of successive cloud layers, but also their presence or absence. 

It is also worth noting that regardless of the value of $\alpha$, H$_2$ is the dominant constituent at the top of the atmosphere in all three cases examined in this section (see column 2 in Fig. \ref{fig:LV}). This contrasts with the results of \citet{ingersoll69}, the foundational first study of runaway greenhouse-induced steam atmospheres,  which demonstrated that, with N$_2$ as the background gas, H$_2$O remains the dominant atmospheric constituent even up to high altitudes and low temperatures if it is the dominant constituent at the planetary surface \citep[see Fig. 2 in][]{ingersoll69}. The abundance of upper-atmospheric H$_2$O in the \citet{ingersoll69} runaway greenhouse simulations leads to efficient H$_2$O photodissociation and subsequent escape, implying efficient planetary desiccation. With hydrogen as the background noncondensable, an upper atmosphere composed mostly of hydrogen is produced by our simulations, such that the H$_2$ may act as a shield, protecting atmospheric H$_2$O from photodissociation. Thus, steam atmospheres may last longer on runaway greenhouse planets with large hydrogen inventories compared to those with large nitrogen inventories, since the planet's water will be destroyed more slowly, though it is also worth noting that the H$_2$ background gas would itself probably escape relatively quickly on these planets.

Clear-sky radiative transfer calculations for the MO atmospheres (Fig. \ref{fig:RT}, left column) demonstrate substantial differences in outgoing radiation for the different $\alpha$ values (subfigure A1) and relatively minor changes to the spectrum of the radiation (subfigure A2). The top-of-atmosphere (TOA) outgoing radiation values are 159, 207, and 335 W m$^{-2}$ for the $\alpha=0.0$, 0.1, and 1.0 cases respectively. So, lapse rate changes from condensate retention can more than double the TOA outgoing radiation from a magma ocean planet, though the inclusion of a stratosphere in radiative balance at the top of the atmosphere would likely reduce the contrast between the $\alpha=1.0$ case and the other two cases, since the upper-atmospheric temperatures in the $\alpha=0.0$ and 0.1 cases are unrealistically cold. Still, this suggests that the lapse rate changes caused by different condensate retention assumptions have substantial influence on TOA outgoing radiation, with potentially important consequences for surface temperature and the timescale of magma ocean crystallization, even without considering the (likely dominant) impact of clouds.

\subsection{Hadean-analogue Exo-Earth after a major Late Veneer impact}
This planetary configuration is meant to represent an analogue to a hot, chemically complex, transiently reducing atmosphere catalyzed by iron delivered to Earth's surface by the disrupted core of a Vesta-sized impactor during the Late Veneer (LV) on Hadean Earth \citep{Genda+2017,Benner+2020,zahnle2020creation}. Impact-induced reducing atmospheres of this type have recently been hypothesized to have played a central role in the origin of life on Earth by allowing for the efficient creation of RNA precursors \citep{Benner+2020,zahnle2020creation}. The specific composition of this atmosphere (see Table \ref{tab:cases}) was drawn from Figure 3 in \citet{zahnle2020creation}, with an additional 500 bars of H$_2$O to represent the $\approx$2 oceans'-worth of water that could have been vaporized by a Vesta-sized impactor \citep[see Table 1 in][]{zahnle2020creation}.  

With full rainout ($\alpha=0$, top row of Fig. \ref{fig:LV}), four species condense at some point in the atmosphere: first H$_2$O near the atmosphere's base, next CO$_2$ at approximately 10$^{-1}$ bar, then CH$_4$ at $\approx$10$^{-4}$ bar, and finally N$_2$ at $\approx10^{-5}$ bar. In actuality, methane and nitrogen would probably not condense because radiative heating would likely warm the upper atmosphere to  $>$100 K, but these calculations are meant to be illustrative. In any event, this sequence of condensations occurs because the lack of retained condensate allows the specific heat to remain at the low levels displayed by gaseous molecules without many internal degrees of freedom ($\approx$30 J K$^{-1}$ mol$^{-1}$), forcing substantial cooling as the gas mixture ascends and expands.

In the $\alpha=0.1$ (middle row of Fig. \ref{fig:LV}) and $\alpha=1.0$ (bottom row of Fig. \ref{fig:LV}) cases, the effective specific heat reaches $\approx$100 J K$^{-1}$ mol$^{-1}$ and $\approx$600 J K$^{-1}$ mol$^{-1}$ respectively as H$_2$O condensates accumulate in the atmosphere. In both of these cases, the high specific heat and correspondingly slow cooling of ascending parcels of air maintains the temperature above the condensation points of CO$_2$, CH$_4$, and N$_2$ throughout the atmosphere. In the $\alpha=1$ case, the cooling with ascent is slow enough that gaseous H$_2$O still makes up $\approx$1$\%$ of the moles in the atmosphere at 10$^{-5}$ bar (subfigure B3), while in the $\alpha=0.1$ case the cooling is strong enough to reduce gaseous H$_2$O to $0.01\%$ of the moles in the atmosphere by 10$^{-2}$ bar (subfigure B2). At pressures lower than about $100$ bar, a majority of moles in the atmosphere of the $\alpha=1$ case are condensed H$_2$O, and below 10 bar, the percentage approaches 99$\%$. 

The LV atmospheres display large differences in the TOA outgoing flux and spectral flux density at the three different $\alpha$ values considered (Fig. \ref{fig:RT}, middle column). The outgoing flux values for the $\alpha=0.0$, 0.1, and 1.0 cases are 38, 192, and 397 W m$^{-2}$ respectively (subfigure B1), an order of magnitude change between the limiting values. However, just like the MO case, it must be noted that a self-consistent stratosphere calculation would markedly reduce this difference. Still, the potential changes in surface climate from changes to outgoing radiation of this magnitude are quite large and may be relevant to the prebiotic chemistry that has been suggested to occur in the aftermath of major Late Veneer impacts. There are also marked differences between the spectra of the LV variants (subfigure B2), with strong absorption lines in the 1-3 micron range present in the $\alpha=1.0$ case but absent in  the $\alpha=0.1$ and 0.0 cases. This is because of the reduced concentration of H$_2$O species in the colder upper atmospheres of the $\alpha=0.0$ and 0.1 cases. 

\subsection{Archean-analogue Exo-Earth near the outer edge of the habitable zone}
Lastly we consider a temperate (290 K) terrestrial planet with an atmosphere consisting of 29 bars of CO$_2$, 1 bar of N$_2$, and 0.01 bar of H$_2$O at the surface (see Table \ref{tab:cases} and Fig. \ref{fig:OHZ}). Here, our Earth-like planet that has passed through violent, rapidly-evolving magma ocean and Late Veneer stages and settled into a stable quasi-equilibrium, with high pCO$_2$ maintained by a carbonate-silicate cycle in response to relatively low instellation \citep[e.g.][]{Walker-Hays-Kasting-1981:negative,graham2020thermodynamic}. This is an example of a hypothetical class of ``habitable'' planet that may exist near the outer edges of the liquid water habitable zones of some stars \citep[e.g.][]{Kopparapu:2013}. With such a high partial pressure, CO$_2$ in this atmosphere acts as a non-dilute condensable component.

This impact of retained condensates on the effective atmospheric specific heat in this configuration is the strongest out of the three case studies. In the top row, where rainout is instantaneous and complete, the atmosphere has a specific heat of between 30 and 40 J K$^{-1}$ mol$^{-1}$ throughout (see subfigure D1). This causes the atmosphere to cool rapidly with height, particularly at pressures less than ~10$^{-1}$ bar where the CO$_2$ has mostly rained out (subfigure B1) and is no longer contributing a significant amount of latent heat. Since we do not account for upper atmospheric radiative heating in the temperature profile, the gas mixture eventually grows so cold from expansion work that N$_2$ begins to condense at around 10$^{-3}$ bar. In actuality, the stratosphere would likely reach a temperature of around 150 K, given the CO$_2$-rich atmosphere and assumption of an outer-habitable zone orbit (i.e. an instellation of something like $40\%$ of Earth's) \citep[e.g.][]{Kopparapu:2013}. This would certainly prevent the N$_2$ condensation in the $\alpha=0$ case. 

Shown in the middle row of Fig. \ref{fig:OHZ}, condensate retention of 10$\%$ leads to nearly an order of magnitude increase in effective specific heat, with the atmosphere reaching $\approx$200 J K$^{-1}$ mol$^{-1}$ at $\approx$10$^{-1}$ bar (subfigure D2). Condensed CO$_2$ makes up a majority of the moles in the atmosphere at pressures lower than 10$^5$ Pa (subfigure B2). The $\alpha=1.0$ case increases the specific heat of the atmosphere by nearly another order of magnitude compared to the $\alpha=0.1$ case, with $c_p>1000$ J K$^{-1}$ mol$^{-1}$ at pressures below 10$^{-1}$ bar (subfigure D3). This further slows the cooling with altitude, allowing CO$_2$'s partial pressure to fall more gradually compared to the N$_2$ partial pressure, such that the gaseous portion of the atmosphere is a mixture of those components in nearly equal parts at pressures lower than approximately 10$^{-1}$ bar (subfigure B3).

The radiative transfer differences for the three $\alpha$ variants of this case are smaller than in the MO and LV cases, though still large enough for potentially significant impacts on surface climate. The TOA outgoing radiation values for $\alpha=0.0$, 0.1, and 1.0 are 81.1, 85.9, and 95.6 W m$^{-2}$. Similar to the MO and LV cases, the extremely cold temperatures reached in the upper atmosphere of the $\alpha=0.0$ variant of this temperate, high-pCO$_2$ atmosphere are unrealistic because radiative balance would almost certainly establish a warmer stratosphere, which would lead to a greater value for TOA outgoing radiation. Still, differences of $\mathcal{O}$(10 W m$^{-2}$) can be significant for surface climate, with the potential to make the difference between a snowball state and a habitable state near the outer edge of the habitable zone. The differences in the spectra between the three $\alpha$ values for the EE case are mostly negligible, though there are order-of-magnitude differences between the $\alpha=0.0$ and 1.0 cases between 6 and 10 $\mu$m.  

\section{Discussion} \label{sec:discussion}

Improving our theoretical capabilities to model exotic atmospheres and compositional inventories of rocky planets is important for interpreting observations of extrasolar planets and further constraining the earliest planetary conditions of the terrestrial planets of the Solar System, for which geochemical proxies are sparse. In this work we derived a novel, self-consistent prescription for the adiabatic lapse rate applicable to multi-component, non-dilute atmospheres on terrestrial worlds. 

As demonstrated in Sect. \ref{sec:discussion}, the changes in lapse rate and thus atmospheric temperature structure can have significant impacts on various aspects of the climate, from the spatial extent of condensing regions to molecular weight stratification, radiating properties, and spectral appearance. The changes for atmospheres with multiple non-dilute condensible species described here may significantly alter the climate during important evolutionary epochs of planetary evolution. Therefore, the methodology presented here will in the future allow a better representation of these atmospheric features, and pave the way to a more complete understanding of planetary evolution. In the following, we will discuss two different, but complementary, aspects of this work in more detail.

\subsection{Implications for the interpretation of exoplanet observations}

Due to the observational bias in exoplanet detections, orbits within the runaway greenhouse limit are currently the dominant host of known rocky planets, such as super-Earths. Eclipse mapping techniques offer clues to the thermal and petrological properties of these extrasolar, terrestrial worlds \citep{2016Natur.532..207D,2019Natur.573...87K} and hold insights into the interaction of atmospheric constituents with underlying, potentially molten mantles. Upcoming ground- and space-based telescopes may enable us to directly image the light radiated from molten \citep{2014ApJ...784...27L,2019A&A...621A.125B} and temperate \citep{2017ApJ...850..121M,2019arXiv190801316Q,LIFE2021a} extrasolar planets on wider orbits. Inverting these observations to infer atmospheric and, eventually, surface properties is sensitive to the underlying model assumptions. Because of the differences in temperature structure and related radiative and compositional properties, the pseudoadiabat derived in this work will enable more detailed reconnaissance of high molecular weight atmospheres and thus the transition from primary to secondary atmospheres \citep{KiteBarnett2020PNAS}. 

In addition, because of the impact on thermal structure, H$_2$O loss and abiotic generation of O$_2$-dominated atmospheres \citep{2015AsBio..15..119L} through upper-atmosphere photolysis and subsequent escape can be affected by variations in retained condensate fraction. Our results here illustrate that variations on condensate retention may affect the H$_2$ abundances in the upper atmosphere, which can alter the degree of photodissociation and thus the lifetime of runaway greenhouse atmospheres on close-in rocky exoplanets. The amount of hydrogen loss in the upper atmosphere can be limited by the coldest region of the atmosphere \citep[the cold trap,][]{2013ApJ...778..154W}. Therefore, in non-dilute, high mean molecular weight atmospheres, deviations in tropopause location may impact the loss rate. Future work shall explore the link between outgassing speciation and retention from primordial evolutionary phases of rocky planets to more realistically capture this feedback between the planetary interior, atmospheric composition, and escape.

\subsection{Implications for Early Earth climate}

A variety of different atmospheric compositions for early Earth have been suggested, in efforts to explain geochemical evidence for surface liquid water during the earliest epochs of Earth's evolution \citep{2001Natur.409..178M,2001Natur.409..175W} despite a faint young Sun, and to reconcile geochemical evidence for an oxidized mantle with the preference \citep{2014AREPS..42..151C,2019Sci...365..903A} of prebiotic chemistry for reduced planetary environments \citep{Benner+2020,2020SciA....6.3419S}. Hydrogen-rich environments have been suggested as promising solutions to both of these questions \citep{2013Sci...339...64W,2020E&PSL.55016546L}, but were likely transient due to the isotopic evidence for limited $p$H$_2$ after the Moon-forming giant impact \citep{2019E&PSL.52615770P}.

In this work we have shown the impact of our parameterization on the lapse rate for both the evolutionary cases of the more conservative CO$_2$-dominated early Earth atmosphere, and the recently suggested, transiently-reducing conditions due to late veneer impacts \citep{Genda+2017,Benner+2020,zahnle2020creation}. These have been suggested to exhibit a variety of complex atmospheric chemistries due to the dominant influence of iron rain-out from the shredded impactor cores. During the following relapse of the atmosphere after such a catastrophic event, the near-surface conditions may intimately depend on the feedbacks between radiative structure and atmospheric chemistry, none of which can be adequately treated with simple single-species temperature structures.

Finally, even earlier, the creation of the primary atmosphere after magma ocean cooldown is impacted by the redox state of the planetary interior, and affects climate and atmospheric spectral appearance \citep[e.g.][]{2020A&A...643A..81K}. The MO cases shown in this work illustrate the impact on the cooling properties expected for such conditions, highlighting the need to appropriately account for multiple condensables under non-dilute conditions, to derive the earliest prebiotic planetary environment of the Hadean Earth.

\subsection{Caveats}
In the evaluation of the case studies presented in this paper, we neglected condensation-inhibited convection, cloud radiative effects, and the formation of a stratosphere which, if included, could significantly impact some of the results. We discuss these phenomena in the following subsections.

\subsubsection{Condensation-inhibited convection}
\citet{guillot1995condensation} and \citet{leconte2017condensation} demonstrated theoretically that the condensation of an atmospheric species with a molar mass higher than that of the non-condensing background gas can produce conditions that stabilize against convective and double-diffusive instability if the condensable is abundant enough. If this convective shutdown occurs where it is optically thick, energy flux through the atmosphere is forced to proceed by radiative diffusion, resulting in sharp superadiabatic temperature gradients in the stabilized regions. This phenomenon can lead to hundreds-of-Kelvin increases in the interior temperatures of solar system gas giants predicted by extrapolating from upper atmospheric observations. 

A number of the atmospheric compositions we consider in this study have condensable components that are of higher molar mass than their dry background components. This is clear when examining vertical stratification of molar mass in Fig. \ref{fig:MO} (subfigures A3, B3), Fig. \ref{fig:LV} (subfigure A3), and Fig. \ref{fig:OHZ} (subfigures A3, B3, C3). In each of those cases, the average molar mass of the atmosphere substantially decreases with height due to the condensation of a heavy, non-dilute atmospheric component, and the concentrations of the condensables exceed the theoretical thresholds for inhibition of convection given by the formula presented in \citet{guillot1995condensation, leconte2017condensation}. Just like in the gassy planet cases explored in the original studies, this inhibition of convection should reduce the efficiency of the vertical transfer of energy through the atmosphere, especially if the gradient in molar mass occurs in an optically thick region. Altering the lapse rate calculation in these regions to account for the dominance of radiative diffusion would lead to larger surface temperatures for a given level of outgoing radiation, in effect enhancing the greenhouse effect of these atmospheres. This effect could be important for the evolution of magma oceans and the climates of planets with thick CO$_2$-N$_2$-H$_2$O atmospheres near the outer edges of liquid water habitable zones.

\subsubsection{Cloud impact on radiative transfer}
In the radiative transfer calculations we carried out for this study, we neglected the impacts of clouds because of the complexities involved in self-consistently determining their properties in one-dimensional, steady-state models \citep[e.g.][]{Kopparapu:2013}. By definition, in all of the cases with $\alpha>0$, clouds would in fact be present,
so our neglect of cloud effects represents a significant simplification. The presence and position of cloud layers in terrestrial planetary atmospheres can have important impacts on both albedo \citep[e.g.][]{yang2014-low,2019Icar..317..583P} and outgoing radiation \citep[e.g.][]{goldblatt2011clouds,kitzmann2016revisiting}, in turn strongly affecting surface climate and the spectral signatures of atmospheres. With an assumed value for $\alpha$, the only additional information required to compute clouds' radiative impacts is the particle size distribution. Future work will assess the feedback between the radiative effects of clouds and the changes in lapse rate predicted by the formula derived in this work.

\subsubsection{Formation of a stratosphere}
In the case studies, we neglected the formation of a stratosphere in radiative balance at the top of the atmosphere, instead focusing solely on the temperature-pressure profiles produced by the pseudoadiabat formula we derived. This leads to very cold temperatures at low pressures, particularly in the cases with $\alpha=0.0$ or 0.1. In actuality, absorption of stellar radiation at the top of the atmosphere would likely cause the atmosphere to display approximately isothermal or even increasing temperatures with height at low pressures, with details like the tropopause height and the cold trap temperature depending on the TOA instellation, the spectrum of radiation received from the star, and the molecular species present. Under conditions with low instellation and an atmosphere with abundant narrow-band IR emitters like CO$_2$, frigid stratospheric temperatures are plausible. Higher instellations or different compositions could lead to a warmer upper atmosphere, which would in some cases prevent condensation of species that condense when the atmosphere is modeled using only the pseudoadiabat. It would also increase outgoing radiation by allowing the upper atmosphere to emit larger quantities of thermal radiation, though the impact of this on surface climate should be modest \citep{Pierrehumbert:2010-book}. Still, without a formula that explicitly accounts for multiple non-dilute condensing components, the influence of stratosphere formation on outgoing radiation and condensation in the upper atmosphere could not be self-consistently evaluated.
\section{Conclusions} \label{sec:conclusions}
In this article, we derive a moist pseudoadiabatic lapse rate formula that allows for any number of condensables, each with arbitrary diluteness and arbitrary retained condensate fraction, and we apply the formula to a set of test cases to illustrate the behavior of terrestrial atmospheres with multiple condensable components and variable retained condensate fraction. The derivation is a generalization of that presented for the single-condensable case in Section 2.6.2 of \citet{Pierrehumbert:2010-book}, and the formula is an expansion on the work of \citet{li2018moist}, which assumed full condensate retention. The introduction of an arbitrary retained condensate fraction is mainly useful for the modeling of atmospheres with non-dilute condensable components, as a large fraction of a non-dilute atmosphere's substance can be converted to solid or liquid condensate. The limit of complete condensate retention is probably unrealistic for atmospheres like this because of precipitation formation and dynamic instabilities from the negative buoyancy of condensate-rich layers. Applying the formula, we find:
\begin{itemize}
    \item Accounting for multiple condensing components with arbitrary retained condensate fraction produces an equation that is more complex than a simple linear superposition of single-component moist (pseudo)adiabats.
    \item When condensate is retained in the non-dilute limit, the effective specific heat per gaseous mole of an atmosphere can increase from tens of J K$^{-1}$ mol$^{-1}$ to hundreds or even thousands of J K$^{-1}$ mol$^{-1}$, strongly influencing the atmospheric lapse rate.
    \item Changes to the atmospheric lapse rate can influence the position and presence of successive cloud layers in atmospheres with multiple condensables, the magnitude of the greenhouse effect, and the pressure at the base of the stratosphere.
    \item The changes to condensate retention cause order-of-magnitude changes to the spectral flux density of outgoing radiation at a variety of wavelengths in all three atmospheric case studies.
    \item In runaway greenhouse atmospheres where hydrogen acts as the background noncondensable, an upper atmosphere composed mostly of hydrogen may be produced. This suggests that H$_2$ can act as a shield to protect atmospheric H$_2$O from photodissociation and hence prolong steam atmosphere lifetimes and increase water retention.
    \item For young rocky planets in the magma ocean phase or during Hadean-like early bombardment epochs, the spectral appearance in the near- to mid-infrared region can be affected by several orders of magnitude when varying the condensate retention fraction, though this ignores the undoubtedly significant impact of clouds on the spectrum.
\end{itemize}
The pseudoadiabat formula derived above enables more complete climate models to account for the effects of varying condensate retention in non-dilute multicondensable atmospheres. Upcoming observations may make use of the newly derived formulation to calculate the detailed radiative-convective balance and infer near-surface properties of extrasolar rocky planets.

\acknowledgements{\textbf{Acknowledgements:}  RJG acknowledges scholarship funding from the Clarendon Fund and Jesus College, Oxford. TL was supported by a grant from the Simons Foundation (SCOL award \#611576). RTP and RB were supported by European Research Council Advanced Grant EXOCONDENSE ($\#$740963). This paper was nucleated by a discussion in a meeting of the Planetary Climate Dynamics group at Oxford. We thank Sarah Rugheimer for useful discussions during the course of the writing of  the paper.}

\bibliographystyle{aasjournal.bst}
\bibliography{biblio3,references}



\end{document}